\documentstyle[graphicx,axodraw,12pt,psfig]{article}
\makeatletter
\def\@cite#1#2{{[{#1}]\if@tempswa\typeout
{IJCGA warning: optional citation argument
ignored: `#2'} \fi}}

\newcount\@tempcntc
\def\@citex[#1]#2{\if@filesw\immediate\write\@auxout{\string\citation{#2}}\fi
  \@tempcnta\z@\@tempcntb\m@ne\def\@citea{}\@cite{\@for\@citeb:=#2\do
    {\@ifundefined
       {b@\@citeb}{\@citeo\@tempcntb\m@ne\@citea\def\@citea{,}{\bf ?}\@warning
       {Citation `\@citeb' on page \thepage \space undefined}}%
    {\setbox\z@\hbox{\global\@tempcntc0\csname b@\@citeb\endcsname\relax}%
     \ifnum\@tempcntc=\z@ \@citeo\@tempcntb\m@ne
       \@citea\def\@citea{,}\hbox{\csname b@\@citeb\endcsname}%
     \else
      \advance\@tempcntb\@ne
      \ifnum\@tempcntb=\@tempcntc
      \else\advance\@tempcntb\m@ne\@citeo
      \@tempcnta\@tempcntc\@tempcntb\@tempcntc\fi\fi}}\@citeo}{#1}}
\def\@citeo{\ifnum\@tempcnta>\@tempcntb\else\@citea\def\@citea{,}%
  \ifnum\@tempcnta=\@tempcntb\the\@tempcnta\else
   {\advance\@tempcnta\@ne\ifnum\@tempcnta=\@tempcntb \else
\def\@citea{--}\fi
    \advance\@tempcnta\m@ne\the\@tempcnta\@citea\the\@tempcntb}\fi\fi}
\makeatother

\newcommand{\nn}{\nonumber}

\newcommand{\gsim}{\lower.7ex\hbox{$\;\stackrel{\textstyle>}{\sim}\;$}}
\newcommand{\lsim}{\lower.7ex\hbox{$\;\stackrel{\textstyle<}{\sim}\;$}}
\newcommand{\be}{\begin{equation}}
\newcommand{\ee}{\end{equation}}
\newcommand{\bea}{\begin{eqnarray}}
\newcommand{\eea}{\end{eqnarray}}

\def\baselinestretch{1}
\begin{document}
\catcode`@=11
\newtoks\@stequation
\def\subequations{\refstepcounter{equation}%
\edef\@savedequation{\the\c@equation}%
  \@stequation=\expandafter{\theequation}
  \edef\@savedtheequation{\the\@stequation}
  \edef\oldtheequation{\theequation}%
  \setcounter{equation}{0}%
  \def\theequation{\oldtheequation\alph{equation}}}
\def\endsubequations{\setcounter{equation}{\@savedequation}%
  \@stequation=\expandafter{\@savedtheequation}%
  \edef\theequation{\the\@stequation}\global\@ignoretrue

\noindent}
\catcode`@=12
\begin{titlepage}

\title{{\bf A Supersymmetric Standard Model of Inflation with Extra
Dimensions}} 
\vskip2in
\author{
{\bf M. Bastero-Gil$^{1}$\footnote{\baselineskip=16pt E-mail: {\tt
mbg20@pact.cpes.susx.ac.uk}}},
{\bf V. Di Clemente$^{2}$\footnote{\baselineskip=16pt E-mail: {\tt
vicente@hep.phys.soton.ac.uk}}}
and
{\bf S. F. King $^{2}$\footnote{\baselineskip=16pt E-mail: {\tt
sfk@hep.phys.soton.ac.uk}}}
\hspace{3cm}\\
 $^{1}$~{\small Centre of Theoretical Physics, University of Sussex,} \\
{\small Falmer, Brighton, BN1 9QJ, U.K.}
\hspace{0.3cm}\\
 $^{2}$~{\small Department of Physics and Astronomy, University of Southampton,} \\
{\small Highfield, Southampton, SO17 1BJ, U.K.}
\hspace{0.3cm}\\
}
\date{}
\maketitle
\def\baselinestretch{1.15}
\begin{abstract}
\noindent
We embed the supersymmetric standard model of hybrid inflation 
based on the next-to-minimal superpotential term $\lambda NH_uH_d$
supplemented by an inflaton term $\kappa \phi N^2$, 
into an extra-dimensional framework, in which all the Higgs fields
and singlets live in the bulk, while all the matter fields live on the brane.
All the parameters of the effective 4d model
can then be naturally understood in terms of a
fundamental (``string'') scale $M_*\sim 10^{13}$ GeV and a
brane supersymmetry breaking scale $10^{8}$ GeV,
of the same order as the height of the inflaton potential during inflation.
In particular the very small Yukawa couplings 
$\lambda \sim \kappa \sim 10^{-10}$ necessary for the model to solve
the strong CP problem and generate the correct effective $\mu$ term
after inflation, can be naturally understood in terms of volume
suppression factors. The brane scalar masses are naturally of order a
TeV while the bulk inflaton mass is naturally in the MeV range 
sufficient to satisfy the slow roll constraints.
Curvature perturbations are generated after inflation
from the isocurvature perturbations of the supersymmetric Higgs as discussed
in a companion paper.

\end{abstract}

\thispagestyle{empty}
\vspace{0.5cm}
\today
\leftline{}

\vskip-19.cm
\rightline{}
\rightline{SUSX-TH 02-026}
\rightline{SHEP 02-30}
\rightline{hep-ph/0211012}
\vskip3in

\end{titlepage}


\setcounter{footnote}{0} \setcounter{page}{1}
\newpage
\baselineskip=20pt

\noindent

\section{Introduction}

Although inflation provides a solution to the flatness and horizon
problems, and is also supported by mounting evidence from detailed
studies of the CMB spectrum \cite{Wang:2001gy},
its relation to particle physics remains obscure. 
Some time ago two of us (BGK) proposed a supersymmetric hybrid inflation
model based on the next-to-minimal superpotential term $\lambda NH_uH_d$
supplemented by an inflaton term $\kappa \phi N^2$
\cite{Bastero-Gil:1997vn}. The motivation for this model was to 
construct a realistic model of inflation which was motivated by 
particle physics considerations. The idea was that, at the end of
inflation, the singlet $N$ would develop a vacuum expectation value
(vev) of order $10^{13}$ GeV, breaking a Peccei-Quinn
symmetry in the process and providing a solution to the
strong CP problem, as well as providing an effective origin
for the Higgs mass term ($\mu$ term) at the TeV scale.
Unfortunately the BGK model appears to suffer from a number
of naturalness problems. The first problem is that in order
to generate a TeV scale effective $\mu$ term, and satisfy
other requirements of inflation, the dimensionless couplings must be
very small $\lambda \sim \kappa \sim 10^{-10}$.
The second problem is that in order for the inflaton to provide curvature
perturbations of the correct order of magnitude the inflaton $\phi$
mass also has to be extremely small being in the eV range.
Finally the height of the inflaton potential of order $10^8$ GeV
is much smaller than the generic $10^{11}$ GeV which is typical
of supergravity explanations for the generation of TeV scale soft
masses~\footnote{In Supergravity the soft masses are given by $m \approx
F_s/m_p$, where $m_p$ is the Planck scale and $\sqrt{F_s}\approx 10^{11}$
GeV is the supersymmetric breaking scale. On the other hand, $F_s^2$ is the
natural order of magnitude for the vacuum energy $V(0)$ (the height of the
inflaton potential).}.  

Recently in \cite{Dimopoulos:2002kt} it was pointed out that 
the second problem of the BGK model, namely that of the eV
inflaton mass, could be alleviated by relaxing the requirement that 
the inflaton be responsible for generating the observed 
curvature perturbations \cite{Wang:2001gy}. 
The basic idea is that the inflaton is only required to 
satisfy the slow roll conditions for inflation, and the curvature
perturbations may be generated after inflation from the isocurvature
perturbations of some late decaying scalar field called the
``curvaton''  \cite{curvaton,others,others2}. In the BGK model
it was pointed out \cite{Dimopoulos:2002kt}
that this means that the inflaton need only
have a mass of order MeV and not eV as in the original version
of the model, thereby alleviating extreme fine-tuning in this model.
However no candidate was proposed for the curvaton, and the remaining
naturalness problems of the smallness of the couplings $\lambda,
\kappa$, the small height of the inflaton potential and the less
extreme but still unnatural requirement of an MeV inflaton mass
was not addressed in \cite{Dimopoulos:2002kt}.
In a companion paper \cite{companion} we show that the Higgs
scalars $H_u,H_d$ of the BGK model could be responsible for generating 
the curvature perturbations responsible for large scale structure.
This represents an alternative to the late decaying scalar
mechanism in which the curvature perturbations are generated
during the reheating stage. 
As in the curvaton approach \cite{Dimopoulos:2002kt} this
allows the inflaton mass to be in the MeV range,
but does not solve any of the remaining naturalness problems
of the model. 

The purpose of the present paper is to show how,
by embedding the BGK model into an extra dimensional framework,
all the remaining naturalness problems of the model may be resolved.
The extra dimensional set-up has all the Higgs fields
and singlets in the bulk, and all the matter fields live on the branes.
All the parameters of the effective 4d model
can then be naturally understood in terms of a
fundamental (``string'') scale $M_*\sim 10^{13}$ GeV and a
brane supersymmetry (SUSY) breaking $F$-term of order $10^{8}$ GeV.
In particular the very small Yukawa couplings 
$\lambda \sim \kappa \sim 10^{-10}$ necessary for the model to solve
the strong CP problem and generate the correct effective $\mu$ term
after inflation, can be naturally understood in terms of volume
suppression factors \cite{kanti,mohapatra}. Also MeV inflaton masses
for scalars in the bulk, and TeV scale soft masses for scalars on the
branes are naturally generated.

The layout of the remainder of the paper is the following. 
In section 2 we review the BGK model in more detail.
In section 3 we embed the model in an extra dimensional framework,
and show how this leads to volume suppressed 4d effective Yukawa
couplings $\lambda, \kappa$ of order $10^{-10}$.
In section 4 we describe the SUSY breaking mechanism due to 
a brane singlet F-term $F_S$, and show how this leads to both TeV scale
soft masses for brane scalars and trilinears
and MeV scale soft masses for bulk scalars such as the inflaton.
Summary and conclusions are presented in section 5.

\section{Brief Review of the BGK Model} 

In this section we first revisit the main features of the 4-dimensional
supersymmetric hybrid model, based on the superpotential\footnote{In
this paper 
superpotential in Eq. (\ref{wnmssm}) is regarded only as the {\it
effective} 4-dimensional superpotential obtained after dimensional reduction.
It has been pointed out in the Ref.~\cite{binetruy}
that in the brane world setup the Hubble parameter $H$ is proportional
to the energy density on the brane, $\rho$, instead of the usual
$H\sim\sqrt{\rho}$ of the standard big bang cosmology. However putting
fields in the bulk whose density dominates over the brane density and
requiring stability of the extra dimensions during the
inflationary period it is possible to show that the standard cosmology is 
recovered~\cite{mohapatra0}. Therefore the 4d results in this section 
remain effectively valid when the theory is embedded in extra
dimensions as is done in the next section.} 
\cite{Bastero-Gil:1997vn} 
\be
W= \lambda N H_u H_d - \kappa \phi N^2 \,,
\label{wnmssm}
\ee
where $N$ and $\phi$ are singlet fields, and $H_{u,d}$ the Higgs fields of
the MSSM.
The first term is familiar from the NMSSM, and the
second terms includes another singlet $\phi$ (the inflaton).
As in the NMSSM,  
the combination $\lambda \langle N \rangle $ gives
rise to an effective $\mu$ term in the Higgs superpotential. 
The usual cubic term of the NMSSM
$N^3$ has been replaced here by an interaction term between $N$ and $\phi$. 
In order to keep the superpotential linear in the inflaton field
$\phi$, other cubic terms in the superpotential are forbidden by imposing a
global $U(1)_{PQ}$ Peccei-Quinn symmetry. 
The global symmetry is broken by the vevs of the singlets, 
leading to a very light axion and solving the strong CP problem
\cite{axions}.  
The axion scale $f_a$ is then set by the vevs of the singlets, and
is constrained by astrophysical and cosmological observations to be
roughly in the window  $10^{10} \,{\rm GeV} \le f_a \le 10^{13}$ GeV 
\cite{axionbounds2,axionbounds3}.  

Inflation takes place below the SUSY breaking scale. Including the
soft SUSY breaking terms, trilinears $A_\kappa$ and masses for the
singlets $m_\phi$, $m_N$, the inflationary potential is given by
\cite{Bastero-Gil:1997vn}: 
\be
V(\phi,N)= V(0) + \frac{\kappa^2}{4} N^4 + ( \kappa^2 \phi^2 - \frac
{1}{\sqrt{2}} \kappa A_\kappa \phi + \frac{1}{2} m^2_N ) N^2 +
\frac{1}{2}m^2_\phi \phi^2 \,,
\label{pot}
\ee
where $\phi$ and $N$ represent the real part of the complex fields,
and we have set the axionic part and the Higgs fields to zero for
simplicity. In addition, we have 
introduced a constant term $V(0)$, whose origin  
will be discussed later. The inflationary trajectory is obtained 
when the inflaton field $\phi$ takes values larger than the critical one
\be  
\phi_c \simeq \frac{A_\kappa}{\sqrt{2} \kappa} \,. 
\label{phic} 
\ee
As long as $\phi > \phi_c$, the $N$ field dependent squared mass is
positive and then $N$ is trapped at the origin; the potential energy
in Eq. (\ref{pot})  is then dominated by the vacuum energy $V(0)$.  
When $\phi$ reaches the critical 
value $\phi_c$, the squared mass of the $N$ field changes sign, and both
fields roll down towards the global minimum at $\phi_0= \phi_c/2$,
$N_0= \phi_c/\sqrt{2}$, ending inflation. 

The required values of the couplings and masses are derived by
combining cosmological and particle physics constraints. 
In order to have slow-roll inflation in the first place, the inflaton mass
$m_\phi$ needs to be small enough compared to the Hubble rate of
expansion $H$, as given by the $\eta$ parameter
\bea
\eta_\phi &= & \frac{\left| m^2_\phi \right |}{ 3 H^2}= 
m_P^2\frac{\left| m^2_\phi \right |}{V(0)} < 1 \,,
\label{eta} 
%
\eea
where $m_P = M_P/\sqrt{8\pi}=2.4\times 10^{18}$ GeV is the
reduced Planck mass, and $\eta_\phi$ is evaluated some
$N$ e-folds before the end of inflation.  
Assuming that $m_\phi$ fulfils the above condition, the other physical
scales in the problem are the soft 
breaking term $A_\kappa \simeq 1 $ TeV, and the axion scale $f_a
\sim \phi_c \sim \phi_0\sim N_0 \sim 10^{13}$ GeV. From
Eq. (\ref{phic}), this unavoidably leads 
to a tiny coupling constant of the order $\kappa \sim 10^{-10}$. The same
applies to $\lambda$, with $\mu=\lambda N_0 \sim 1$ 
TeV. Thus, demanding a zero vacuum energy at the global minimum,
$V(\phi_0,N_0)=0$, the height of the potential during inflation is 
given by,
\be
V(0)^{1/4} \simeq \sqrt{\frac{\kappa}{2}} N_0 \simeq (10^8 \, {\rm GeV}) \,. 
\label{v0}
\ee
The Hubble parameter during inflation is then of the order of
$O$(10 MeV). And from Eq. (\ref{eta}),   
this means that inflaton soft mass $m_\phi$ can be at most of the order of
some MeVs in order to satisfy the slow roll conditions.

In order to meet the COBE value $\delta_H= 1.95 \times 10^{-5}$
\cite{cobe}, we would require having $\kappa m_\phi \sim 10^{-18}$
 GeV, i.e., a tiny inflaton mass  of the order of a few 
\footnote{We notice that this value   
can be entirely due to 1-loop radiative corrections  $\delta m_\phi^2
\sim \kappa^2 (\kappa \phi_c)^2$, once the 
tree-level value is set to zero.} eV. This is a much stronger
constraint on  the mass of the inflaton than just requiring the
inflaton to be a ``light'' field during inflation and satisfying the
slow-roll condition Eq. (\ref{eta}). 

As we discuss in a companion paper \cite{companion}
primordial curvature perturbations can be originated by the
Higgs perturbations instead of the inflaton perturbations. The inflaton mass
is then only restricted by the slow-roll condition, 
and therefore no extremely tiny values of the masses are
required. However, the spectral index of the spectrum of curvature
perturbations  is now controlled by the Higgs parameters:
\be
n-1 \simeq 2 m_h^2/(3 H^2) \,,
\ee
which is constrained by observations \cite{Wang:2001gy}
to be $n< 1.06$. And this
is a slightly stronger constraint that just demanding slow-roll, i.e.,
for the Higgs mass we will have $m_h < 0.3 H\simeq 3$ MeV.  

\section{Embedding the BGK model in extra dimensions}

We now embed the BGK model into an extra-dimensional framework, 
in which all the Higgs fields
and singlets live in the bulk, while all the matter fields live on the brane.
In this section we shall show how the very small Yukawa couplings 
$\lambda \sim \kappa \sim 10^{-10}$ necessary for the model to solve
the strong CP problem and generate the correct effective $\mu$ term
after inflation, will be naturally understood in terms of a volume
suppression factor $(M_*/m_p)^2$ where $M_*$ is a
fundamental (``string'') scale $M_*\sim 10^{13}$ GeV and
$m_p$ is the effective reduced Planck scale whose value will also
be explained. In the next section we shall consider SUSY breaking
and show how the required MeV soft masses for the inflaton in the 
bulk and the TeV soft masses on the brane may result from a
brane supersymmetry breaking $F$-term of order $10^{8}$ GeV,
which also naturally sets the scale for the height of the inflaton
potential $V(0)$.

\begin{figure}[h]   
 \begin{center}  
  \begin{picture}(420,190)(0,0)
   \Line( 100, 110 )( 145, 185 )
   \Line( 100,  15 )( 145,  90 )
   \Line( 100,  15 )( 100, 110 )
   \Line( 145,  90 )( 145, 185 )
   \Text( 100,   5 )[c]{$y_j=0$}
   \Text(  90, 110 )[r]{``Yukawa Brane''}
   \Text(  90,  90 )[r]{$Q_i, U_i, D_i$}
  
   \Text(210,130)[]{{\large $\hat G_A^{(1)} , \,\, \hat \phi, \,\, \hat N$}}
   \Text(210,100)[]{{\large $\hat H_u, \,\, \hat H_d$}}
   \Text(200,70)[]{{\large $4+d$-Dim.}}
   \Line( 255, 110 )( 300, 185 ) 
   \Line( 255, 15 )( 300, 90 )
   \Line( 255, 15 )( 255, 110 ) 
   \Line( 300, 90 )( 300, 185 )
   \Text( 255, 5 )[c]{$y_j = \pi R$}  
   \Text( 310, 90 )[l]{SUSY Breaking Brane} 
   \Text( 310, 70 )[l]{gauge singlet $S$, $\hat G_B^{(2)}$} 
  \end{picture}
 \end{center} 
  \caption{{\small The model showing the parallel 3-branes spatially 
separated along $d$ extra dimensions with coordinates ${\bf y} =
   (y_1,\dots, y_d)$ and a common radius $R$.}} 
\label{prev}
\end{figure}
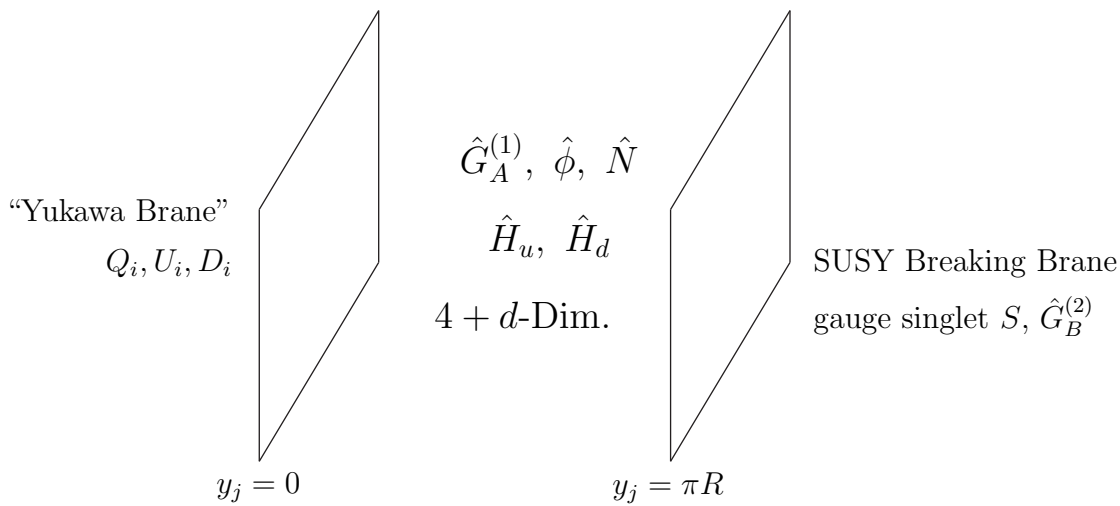

Let us consider two 3-branes spatially separated along $d$ extra
dimensions with a common radius 
$R$. These extra dimensions are compactified on some orbifold that
leads at least to two fixed points  
at $y_j=0, \pi R$ ($j=1,\dots, d$), where the two D3-branes are located.  
All the quarks/squarks fields ($Q_i, U_i, D_i$, where
$i=1,2,3$) live on the ``Yukawa''  
brane at\footnote{In this brane, for 
reasons we shall discuss later, we define the quark's Yukawa couplings. 
In the section 4.4 we shall address the issue regarding the localization
of leptons (so far they can live either in the 
``Yukawa'' brane or in the SUSY breaking brane).} $y_j=0$,  
while SUSY is broken by the F-term of a gauge singlet field S on the
SUSY Breaking Brane at $y_j=\pi R$. 
The gauge fields $\hat G_A^{(1)}$, 
the inflaton field $\hat\phi$, the singlet field $\hat N$ and both 
higgses $\hat H_u$ and $\hat H_d$ feel all the dimensions of the
theory ($4+d$-dimensions). Also we include an additional gauge group
$\hat G_B^{(2)}$ in such 
a way that at some scale $M$ the total gauge group\footnote{In this paper we are not going to considered or
specify a particular gauge group for $G_A^{(1)}\times G_B^{(2)}$, it
can be either some string motivated gauge group
(i.e Pati-Salam group, $E_8$, etc) or some GUT group ($SU(5), SO(10)$,
etc).} 
$G_A^{(1)}\times G_B^{(2)}$ breaks down to 
the Stantard Model gauge group $G_{SM} = SU(3)\times SU(2)\times
U(1)$.  This is depicted in 
Fig.~\ref{prev}.     

In each of the fixed points of the full manifold we have $N=1$
supersymmetry. Nevertheless, in order to get supermultiplets associated
with the Kaluza-Klein (KK) tower after compactifying the $4+d$
dimensions down to 4 dimensions, we need that the infinite degrees of
freedom to fall down to 
some extend supersymmetry.  Strictly speaking, the extra Kaluza-Klein
tower of states will effectively be $N=2$ supersymmetric only for
one or two extra dimensions. For higher values of $d$, the situation
is a bit more complicated. For example, for $d=6$ we naively expect the
Kaluza-Klein tower of states to be $N=4$
supersymmetric~\cite{lykken}.  In general, the enhanced supersymmetry
for the excited Kaluza-Klein arises because the minimum number of
supersymmetries in higher dimensions (as counted in terms of
four-dimensional gaugino spinors) grows with the space-time
dimensions. However, by making suitable choices of orbifolds, it is
always possible to project the relevant Kaluza-Klein towers down to
representations of $N=2$ supersymmetry, even if
$d>2$~\cite{dienes}. Hence, without loss of generality, we shall
consider $N=2$ supersymmetric Kaluza-Klein towers for arbitrary values
of $d$.

In a $ N = 2$ supersymmetric theory there is a global $SU(2)_R$
automorphism group defined in the supersymmetric algebra. The
off-shell hypermultiplet $\Phi_a$ is given by\footnote{For more
detail about the off-shell formulation of the vector and
hypermultiplets in $N=2$ supersymmetry see for example  
Ref.~\cite{peskin}.} 
$\Phi_a = (\phi_a^l, \Psi_a, F_a^l)$,
where $l=1,2$ is the $SU(2)_R$ index and $\Psi_a =
(\psi_{a,L}, \psi_{a,R})$ is a Dirac spinor. On the other hand, it is
well known that in $N = 2$ there is no $SU(2)_R$ invariant cubic
interactions involving 
hypermultiplets $\Phi_a$~\cite{sohnius}.  One possible way to define the
supersymmetric Yukawa couplings is sticking the superpotential in one
of the fixed points ($D3$-branes) of the orbifold, where only one of
the supersymmetry $N=1$ survives after the orbifolding. This is what
we have called the Yukawa brane.

The Lagrangian associated with the superpotential can be written
as\footnote{For simplicity, we set all the Yukawa couplings
except the third generation ones to zero.},
\begin{equation}
{\cal L}^W_{4+d} = - \int \, d^2\theta \; \delta^d({\bf y})
\left[\frac{\hat\lambda}{M_*^{\frac{3d}{2}}}\hat N \hat H_u \hat H_d 
-\frac{\hat\kappa}{M_*^{\frac{3d}{2}}} \hat\phi
\hat N^2 
+\frac{\hat y_t}{M_*^{\frac{d}{2}}} Q_3\hat H_u U_3 
+\frac{\hat y_b}{M_*^{\frac{d}{2}}} Q_3\hat H_d D_3 \right]\,,
\label{lagran1}
\end{equation}
where the couplings and fields with a hat mean couplings and fields
in extra dimensions, being $\lambda$ and $\kappa$ the couplings defined in
Eq.~(\ref{wnmssm}), and $y_t$ ($y_b$) is the top (bottom) Yukawa coupling.
$\delta^d({\bf y})$ is the d-dimensional generalisation of the delta
function. $M_*$ is the string scale in $4+d$ dimensions and it is related 
to the Planck scale in four dimensions through the very well known formula:
\be
m_p^2 = M_*^{2+d} V_d \,,
\label{planck}
\ee
with $V_d\approx R^d$ the volume factor of the compact manifold
defined in the extra dimensional bulk.  The bulk fields have a mass
dimensions $1+d/2$, while the brane fields have the standard  mass
dimension 1. The higher dimensional fields lead to
non-renormalisable interaction terms, where the suppression is now 
given by the fundamental scale in $4+d$ dimensions ($M_*$) instead of
the four dimensional Planck mass ($m_p$).

As we have explained above, the full theory should be written in terms
of hypermultiplets and vector multiplets in $N=2$
supersymmetry. However, the Lagrangian (\ref{lagran1}) is a function
of the $N=1$ supermultiplets ($\phi^1, \psi_L,F^1$) only, because we are 
assuming that the orbifolding casts the two
$N=1$ multiplets in the $N=2$  hypermultiplet in such a way that just
one of them is even under some orbifold discrete group (for example $Z_2$)
and then at the fixed point only this multiplet is different from
zero\footnote{Notice that terms like $\hat\phi (\partial_y \hat
N')^2$, where $N'$ is the supermultiplet which belongs to the {\it other}
$N=1$ supersymmetry, can be allowed by the orbifold symmetry in the
Lagrangian (\ref{lagran1}), but are heavily suppressed by higher
powers of the string scale $M_*$, and can therefore be neglected.}.

Upon dimensional reduction a plethora of particles (KK-modes)
comes out in the effective four dimensional Lagrangian. In fact,
there are mixing among different KK-numbers since the interaction
(\ref{lagran1}) does not preserve the translational invariance along
the extra dimensions due to the presence of the delta function
($\delta^d({\bf y})$) which breaks explicitly the Poincare invariance
of the theory. However making some assumption about the number of
extra dimensions, $d$, we can neglect the contribution from the infinite
tower of KK and write down the effective lagrangian as function only of the
zero mode of each bulk fields. Indeed, using Eq.~(\ref{planck}) we have 
that the compactification scale $1/R$ is given by
\be
\frac{1}{R} = M_* \left(\frac{M_*}{m_p}\right)^{2/d}\,.
\ee
On the other hand, we will see that consistency of the model demands 
the string scale of the theory be present at some intermediate
scale; in particular it has to be of the same order of the axion scale 
$M_* \sim f_a \sim 10^{13}$ GeV. This imply that if the number of extra
dimensions is larger than two, $d>2$, the compactification scale
(using Eq.~(\ref{v0})) is then $1/R>V(0) \sim 10^{8}$ GeV.
   
This means that the energy scale for inflation (governed mainly by the
vacuum energy) is below the first excitation of the KK propagating in
the bulk. Therefore, the KK-modes are not produced during the early
stage of inflation and then the decoupling of these particles 
is a good approximation for our purpose. From now on, we only consider
a number of extra dimensions larger than two, even if the result
we will show here will be independent of the number of extra dimensions.

We are going now to study in details the effective four dimensional yukawa 
couplings and the gauge coupling.

\subsection{Yukawa Couplings}
After integrating out the $d$ extra dimensions and considering only the
zero mode of the bulk fields, from Eq.~(\ref{lagran1}) we get
\bea
{\cal L}_{4}^{W} &=& -\int\, d^2\theta \,
\left[  
\hat\lambda \left(\frac{M_*}{m_p}\right)^3 N H_u H_d 
-\hat\kappa\left(\frac{M_*}{m_p}\right)^3 \phi N^2 
+\hat y_t \left(\frac{M_*}{m_p} \right) Q_3 H_u U_3 \right.\nonumber \\
&+& \left.\hat y_b \left(\frac{M_*}{m_p} \right) Q_3 H_d D_3\right] \,, 
\label{lagran2}
\eea
where now all the fields are four dimensional ones. From the last
equation we found that the four dimensional couplings are naturally
suppressed if $M_* < m_p$,
\bea
\lambda &=& \left(\frac{M_*}{m_p}\right)^3 \hat\lambda, \qquad \qquad   
\kappa = \left(\frac{M_*}{m_p}\right)^3 \hat\kappa, \qquad \qquad \nn
\, \\
Y_t = (y_t) &=& \left(\frac{M_*}{m_p}\right) \hat
y_t, \qquad Y_b= (y_b) =
\left(\frac{M_*}{m_p}\right) \hat y_b \,.
\label{couplings}
\eea   
A natural assumption is consider all the multidimensional couplings to
be of the same order,  
\be
\hat\lambda\sim\hat\kappa\sim \hat y_t \sim \hat y_b  \,,
\label{assumption}
\ee
in which case we get the following relationship between the Yukawa
couplings, 
\be
\frac{\lambda}{Y_{t(b)}} \sim \frac{\kappa}{Y_{t(b)}} \sim
\left(\frac{M_*}{m_p}\right)^2 \,. 
\label{bingo}
\ee
Therefore, if $M_* \approx 10^{13}$ GeV we naturally get $\lambda \sim
\kappa \sim {\cal O}(10^{-10})$ where the Yukawa couplings for the
third generation  
are of the order one, $Y_u \sim Y_d \sim {\cal O}(1)$. Notice that the
$4+d$-dimensional couplings  
in Eq. (\ref{couplings}) are extremely large. This only indicates that
our $D$-dimensional model is non-perturbative. The ``duality'' through
dimensional reduction between a non-perturbative theory in extra dimensions
and a perturbative theory in the effective four dimensions has been pointed
out some time ago in Ref.~\cite{antoniadis}. 

\subsection{Gauge Coupling}
The Lagrangian in $4+d$ dimensions associated with the gauge coupling for
the Higgs (bulk fields) and the quarks (brane fields) has the following
form:
\be
{\cal L}_{4+d}^g = \left[ \frac{\hat{g}}{M_*^{d/2}}
\gamma_M\hat{A}^M\bar{Q_i}Q_i + \left(Q_i \leftrightarrow U_i, D_i
\right)\right] \delta^d({\bf y}) + \left[\frac{\hat{g}^2}{M_*^{d}}\hat{A}_M
\hat{A}^M \hat{H}^\dagger_u\hat{H}_u + \left(\hat{H}_u \leftrightarrow
\hat{H}_d\right) \right] \, ,
\ee 
where $\hat{A}_M$ are the gauge boson in higher dimensions, being $M= \mu,
5=0,1,2,3,5$. The bulk gauge fields have a mass dimensions $1+d/2$ (like
the Higges fields)~\footnote{The gauge coupling remains dimensionless.}. 
After integrating out the $d$ extra dimensions from the above Lagrangian we get
\be
{\cal L}_{4}^g = \left[ \frac{\hat{g}}{M_*^{d/2}}
\gamma_\mu\left(\frac{A^\mu}{\sqrt{V_d}}\right)\bar{Q_i}Q_i + \left(Q_i \leftrightarrow U_i, D_i
\right)\right] + \left[\frac{\hat{g}^2}{M_*^{d}}\left(\frac{A_\mu A^\mu}{V_d}\right)
\left(\frac{H^\dagger_u H_u}{V_d}\right) + \left(\hat{H}_u \leftrightarrow
\hat{H}_d\right) \right] \times V_d 
\ee
Notice that the fifth component of the gauge fields, $A_5$, has been removed
from the Lagrangian since it does not have zero mode. From the previous
Lagrangian and using the eq. (8) we can read the effective four dimensional gauge coupling as
\be
g = \left(\frac{M_*}{m_p}\right) \hat{g} \, .
\label{g}
\ee  
Comparing (\ref{couplings}) and (\ref{g}) we observe that both the gauge
coupling and the yukawa couplings in four dimensions have the same
suppression factor. Thus $g/Y_{t(b)}\sim 1$ assuming that their higher
dimensional couplings are of the same order.

\section{Supersymmetry Breaking}

We shall suppose that SUSY is broken by the F-term of a $4D$
gauge-singlet field 
$S$ on the source brane localised at the fixed point $y_p = \{y_i\} =
\pi R$, and mediated across the extra dimensional space by bulk fields
propagating in a loop 
correction like gaugino mediations~\cite{david,vicente}.
Moreover, $S$ is also neutral under the $U(1)_{PQ}$
symmetry. Because of that, no $\mu$-term is generated by the
Giudice-Masiero mechanism in this 
model~\cite{giudice}. The solution to the $\mu$ problem relies on  
the coupling of the singlet $N$ to the Higgses. Like in the NMSSM the
$\mu$-term is given by $\mu \sim \lambda N_0$, once the singlet $N$ gets a
non-zero vev $N_0$ after inflation. All the bulk fields (gaugino, higgsino,
higgs, inflaton, singlet $N$) get a tree-level SUSY mass term through
direct coupling with the SUSY breaking brane. The rest of the particles
which live in the Yukawa brane only get 1-loop SUSY mass terms and then
they will be neglected in this paper. 

In the next subsections we are going to discuss the origin of all the SUSY
breaking terms necessary to generate the inflaton potential
Eq. (\ref{pot}): the vacuum energy $V(0)$, the trilinear $A_k$ term and
the quadratic mass term, $m_N^2$ and  $m_\phi^2$.  
For completeness we will discuss the other soft terms as well,
i.e. the soft mass term for the Higgses, the $B\mu$ terms and the gaugino 
masses. In the SUSY breaking sector there are two free parameters, 
the F-term of the singlet $S$ ($F_S$) and the cutoff $M_*$. However demanding
the solution of the CP-problem, and  imposing that the $F_S^2$ term
explain the origin of the vacuum energy, we will see that all the
parameters (both dimensionful and dimensionless) of the potential Eq.  
(\ref{pot}) are fully determined.

\subsection{Vacuum energy.} 
The SUSY breaking brane will typically introduce a vacuum energy of 
the order of $F_S^2$ providing a vacuum energy $V(0)$ in the 
potential (\ref{pot}). We simply set this constant from the Lagrangian in
$4+d$-dimensions:
\be
\Delta {\cal L}_{4+d}^{soft} = -  \int\, d^4\theta\, \delta^d ({\bf
y} - y_p) S^\dagger S \,.  
\ee
In the effective four dimensions and when SUSY is broken, we get a 
vacuum energy
\be
V(0) = F_S^2 \,.
\label{FS}
\ee
From Eq. (\ref{v0}) we see that the F-term has to be 
$\sqrt{F_S} \sim 10^8$ GeV. This result is indeed interesting because
it states that the vacuum energy and the SUSY breaking scale are of the same
order of magnitude avoiding any fine-tuning regarding the K\"ahler 
potential~\cite{mar}.  

\subsection{Trilinear soft terms for scalars.}
The trilinear soft terms allowed for the PQ charges are
\be
\Delta {\cal L}_{4+d}^{soft} = - \int\, d^2\theta\, \delta^d ({\bf 
y} - y_p) \left(  
\frac{\hat\lambda}{M_*^{\frac{3d}{2}}} \hat N \hat H_u \hat H_d 
-\frac{\hat\kappa}{M_*^{\frac{3d}{2}}} \hat\phi \hat N^2 
\right) \frac{S}{M_*} \,.
\label{trilinear}
\ee
From the second term in Eq.~(\ref{trilinear}) we get the $A_k$ term defined in 
the potential (\ref{pot}), while the $B\mu$-term  arises from the
first term of Eq.~(\ref{trilinear}) when the $N$ field develops a vev
at the end of inflation.
 
Integrating out the extra dimensions coordinates we get
\be
\Delta {\cal L}_{4}^{soft} = - \int\, d^2\theta\, \left(
\lambda N H_u H_d  - \kappa \phi N^2  \right) \frac{S}{M_*} \,,
\label{trilinear1}
\ee
where we have used  Eq.~(\ref{couplings}) to redefine the
effective Yukawa couplings $\lambda$ and $\kappa$.  When the F-term of the
singlet $S$ gets a vevs ($F_S$), from  Eq.~(\ref{trilinear1}) we
obtain an $A_k$-term during inflation and a $B$-term at the end of inflation,
\be
A_{k} \sim B \sim \frac{F_S}{M_*} \,. 
\label{A-term}
\ee

We have seen that $\sqrt{F_S} \sim 10^8 $ GeV, but $M_*$ seems to be a 
free parameter. From the minimisation of the potential (\ref{pot}) we get 
the vev of the field $N$ given by  $N_0 = A_k/(\sqrt{2} k)$. Using
Eqs. (\ref{bingo}) and (\ref{A-term}) we get
\be
N_0 = \frac{m_p^2}{\sqrt{2}}\frac{F_S}{M_*^3} \,.
\label{eq1}
\ee
On the other hand, from Eqs.~(\ref{v0}),(\ref{bingo}),(\ref{FS}) we have
\be
F_S^{1/2} = \left(\frac{M_*}{m_p}\right) \frac{N_0}{\sqrt{2}} \,. 
\label{eq2}
\ee
Casting the last two equations together we can relate 
$M_*$ with $N_0$ as\footnote{Note that $N_0$ is the
effective field in four dimensions and this field can be larger 
than the string scale $M_*$. The point is that the higher dimensional 
field $\Phi_d$ carries a volume suppression factor, $\Phi_d =
\Phi_4/\sqrt{V_d}$, where $\Phi_4$ is four dimensional field and $V_d$ 
is the extra dimension volume. When we use the above
relationship between $\Phi_d$ and $\Phi_4$ and integrate out the extra 
dimensions, and use the relation (\ref{planck}), the natural 
cutoff for the effective four dimensional field is seen to be $m_P$ and
not $M_*$.}
\be
N_0 = \sqrt{8} M_* \,.
\ee
This means that if we want to solve the CP-problem in our model, we 
immediately need a string scale defined at $M_* \sim 10^{13}$ GeV. Using
also that $\sqrt{F_S}\sim 10^8$ GeV, from  Eq.~(\ref{A-term}) we get 
$A_k \sim B \sim 1$ TeV.

\subsection{Quadratic soft terms for bulk scalars.}
The quadratic soft masses for the scalars are given by 
\be
\Delta {\cal L}_{4+d}^{soft} = -  \int\, d^4\theta\, \delta^d ({\bf
y} - y_p) 
\frac{c_X}{M_*^{d}} \hat X^\dagger \hat X
\frac{S^{\dagger}S}{M_*^2}  \,,
\label{soft}
\ee
where $X$ runs over all the bulk fields, $ \phi, N, H_u, H_d$ and 
$c_X$ are constants of the order one. After dimensional reduction, 
we have the $4D$ Lagrangian,
\be
\Delta {\cal L}_{4}^{soft} = - \int\, d^4\theta\, \frac{c_X}{(M_*^d V_d)} 
X^\dagger X \frac {S^{\dagger}S}{M_*^2} \,.    
\label{soft1}
\ee
Using Eq.~(\ref{planck}) we get the  mass
term for the scalars when the F-term of the field $S$ gets a vev: 
\be
m_X^2 = c_X \left(\frac{F_S}{m_p}\right)^2 \,.
\label{mass-term} 
\ee
From Eqs.(\ref{A-term}) and (\ref{mass-term}) we see that the values for
the trilinear  and the mass terms are non equal (non-universality) as long
as the Planck scale in four dimensions, $m_p$, and the Planck scale in $4+d$ 
dimensions, $M_*$, are different. 
In fact, their ratio is given by:
\be
\frac{m_X}{A_\kappa} = \sqrt{c_X}\left(\frac{ M_*}{m_p}\right) \,. 
\ee
Therefore, we have that the quadratic soft term for bulk fields
are small, in particular $m_X/\sqrt{c_X} \ll A_\kappa$. Below we will
see that this is no any more true for particles defined in one of the branes.
For $M_* \sim 10^{13}$ GeV, $\sqrt{F_S}\sim 10^8$ GeV and
imposing\footnote{One might think  
that the operators (\ref{soft}) arises integrating
out some massive string excitation propagating a 1-loop. It turns out that
the coefficient $c_X$ has to contain the 1-loop factor, 
$c_X \approx 1/(4\pi)^2$.}
$c_X \sim 1/(4\pi)^2 \sim {\cal O}(10^{-2})$, from the last equation
we get a very tiny soft  masses for the bulk fields
\be
m_\phi\sim m_N \sim m_{h_u} \sim m_{h_d} \sim {\cal O}(1 \mbox{MeV}) \,.
\ee

The quadratic soft mass for the inflaton, $m_\phi^2$, generated in the
SUSY breaking brane, is the same that 
appears in the inflaton potential (\ref{pot}). As we have already said 
following Eq.~(\ref{eta}), this mass has to be $m_\phi< {\cal O}(1$ MeV) in
order to satisfy the slow-roll condition for the potential. However, 
this mass is quite large to satisfy the COBE constraint. 
Nevertheless in our model the inflaton does not play an important role to 
generate the density perturbation, instead
a new mechanism is proposed in a companion paper~\cite{companion} in which
the Higgs field can generate the large-scale curvature perturbation from
an efficient conversion of isocurvature perturbation to curvature one
during the reheating era.

\subsection{Quadratic soft terms for brane scalars.}

So far we have discussed how to generate the soft terms for the fields living
in the bulk. The situation is slightly different for the
scalars living on one of the branes. 
The scalars living on the SUSY breaking brane get a soft mass term from 
the following operator:
\be
\Delta {\cal L}_{4+d}^{soft} = -  \int\, d^4\theta\, \delta^d ({\bf
y} - y_p) 
c_Y Y^\dagger Y \frac{S^{\dagger}S}{M_*^2}  \,,
\ee
where $Y$ runs over all the brane's superfields. This leads a soft mass term,
\be
m_Y = \sqrt{c_Y}\left(\frac{F_S}{M_*}\right) \,.
\ee
Using again $\sqrt{F_S}\sim 10^8$ GeV and $M_* \sim 10^{13}$ GeV, we got
a soft term, $m_Y \sim 1$ TeV.

On the other hand, the
bulk scalar masses receive a suppression factor $M_*/m_p$ with respect
the former case due to the finite extra dimension volume. In quantum 
field theory one is free to choose in which branes the
particles live. It can be either the SUSY breaking brane or what we
have called the Yukawa brane. 
A possible motivation to put fermions in the 
Yukawa brane is to solve the FCNC problem since all the interactions which
violate flavour will be suppressed by a factor $exp(-M_* R)$ \cite{david}. 
However, at
tree level all their soft masses will be zero since there is no contact term
with the singlet $S$. Therefore the only way to produce soft masses in
this case will be through radiative corrections via renormalisation
group running. In the case of
squarks, the main contribution at 1-loop comes from the gauginos and the
exact spectra will be quite similar to what happen with non-scale
supergravity or gaugino mediation \cite{david}. 
It is well known that there may be 
phenomenological difficulties with such models in the
slepton sector \cite{Schmaltz:2000ei}.
One possibility is
to leave the quarks/squarks on the Yukawa brane but localise the lepton sector
in the SUSY breaking brane. In this way the sleptons will get TeV soft masses
and we still have the FCNC problem for the quark sector resolved.

\subsection{Gaugino mass}

The gauge group in our model is given by the direct product of two
groups, $G_A^{(1)}\times G_B^{(2)}$. This group will diagonally 
break down to the Standard Model gauge group. 
The gaugino mass for the gauge group which live in the bulk, $G_A^{(1)}$, 
is given by the operator
\be
\Delta {\cal L}_{4+d}^{soft} = -  \int\, d^2\theta\, \delta^d ({\bf
y} - y_p) \frac{c_\lambda^{(1)}}{M_*^{d}} \left(W_\alpha^{(1)} W^{\alpha
(1)}\right)\frac{S}{M_*} \,,
\label{gauge1}
\ee 
where $c_\lambda^{(1)}$ is a constant of the order one and $W_\alpha^{(1)}$
is the field strength of the gauge group $G_A^{(1)}$. After dimensional
reduction we get a soft gaugino mass for the group  $G_A^{(1)}$:
\be
m_{\lambda^{(1)}} = \frac{F_S}{M_*}\frac{1}{(M_* R)^d} = \frac{F_S}{M_*}
\left(\frac{M_*}{m_p}\right)^2 \,,
\ee
where we have used the relation Eq. (\ref{planck}) in the last equality. This
mass is very tiny; in fact, using $M_* \sim 10^{13}$ GeV and 
$\sqrt{F_S} \sim 10^8$ GeV, we found $m_{\lambda^{(1)}} \sim {\cal O}(100$
eV). In the case the only group in our model were the Standard
Model one embedded in extra dimensions, this result would rule out this setup
by direct search of the gauginos. 
However, localising other gauge group, $G_B^{(2)}$,
in the the SUSY breaking brane it is possible to overcome this 
problem\footnote{This group can either be a replica of the same bulk gauge
group $G_A^{(1)}$ or a different one.}. 
   
The gauginos of the group $G_B^{(2)}$ get a mass through the operator
\be
\Delta {\cal L}_{4+d}^{soft} = -  \int\, d^2\theta\, \delta^d ({\bf
y} - y_p) c_\lambda^{(2)} \left(W_\alpha^{(2)} W^{\alpha
(2)}\right)\frac{S}{M_*}\,.
\label{gauge2}
\ee 
Using dimensional reduction we obtain,
\be
m_{\lambda^{(2)}} = \frac{F_S}{M_*} \,.
\ee
This mass is exactly the same than that for  the $A_k$-term
(\ref{A-term}), an therefore 
using $M_* \sim 10^{13}$ GeV and $\sqrt{F_S} \sim 10^8$ GeV we get
$m_{\lambda^{(2)}} \approx 1$ TeV. Hence we have that $m_{\lambda^{(2)}}
>> m_{\lambda^{(1)}}$. 

Once the total gauge group is diagonally breakdown to the Standard Model
gauge group, $G_A^{(1)}\times G_B^{(2)} \longrightarrow G^{SM}$, the
eigenvalue mass for the lowest states are given by~\cite{steve}
\be
m_{\lambda^{SM}} = f(g_1,g_2)m_{\lambda^{(1)}} +
h(g_1,g_2)m_{\lambda^{(2)}} \approx m_{\lambda^{(2)}} \approx 1
\mbox{TeV} \,,
\ee
where $f(g_1,g_2)\sim h(g_1,g_2)\sim {\cal O}$(1) are some function
of the large gauge group $G^{(1)}$ and $G^{(2)}$. 
It turns out that the Standard Model gauginos in our model have mass of the
order of the SUSY breaking scale $\sim {\cal O}$(TeV).
 
\section{Summary}

We have shown how,
by embedding the BGK model into an extra dimensional framework,
all the naturalness problems of the model may be resolved.
An underlying assumption of our approach is that the
radii of the extra dimensions are stabilised for example
by the mechanism proposed in \cite{poppitz}.
The extra dimensional set-up has all the Higgs fields
and singlets in the bulk, and all the matter fields live on the branes.
All the parameters of the effective 4d model, including the Planck scale,
can then be naturally understood in terms of a
fundamental ``string'' scale $M_*\sim 10^{13}$ GeV and a
brane SUSY breaking term $\sqrt{F_S}\sim 10^{8}$ GeV.
Once the number of extra dimensions $d$ is specified
the reduced Planck scale $m_p$ in Eq. (\ref{planck}) then fixes the size of
the extra dimensions.
From these scales everything else follows: the height of the inflaton
potential during inflation is of order $\sqrt{F_S}$; 
the singlet vevs after inflation associated with the 
axion solution to the strong CP problem are of order $M_*$;
the small Yukawa couplings $\lambda \sim \kappa \sim 10^{-10}$
necessary for the model to solve
the strong CP problem and generate the correct effective $\mu$ term
after inflation, are of order $(M_*/m_p)^2$; 
the TeV scale soft masses and trilinears 
for scalars on the branes is naturally understood as 
$F_S/M_*$; the MeV inflaton masses for scalars
in the bulk are suppressed relative to the TeV scale soft masses
by a factor $M_*/m_p$. Although 
we do not address the question of neutrino masses in this paper,
we note that the natural scale of our model $M_* \sim 10^{13}$ GeV is
also typical of right-handed Majorana masses in see-saw models.

\section*{Acknowledgements}
We wish to thank David Lyth for useful discussions.
V.D.C and S.F.K. would like to thank PPARC for a Research Associateship
and a Senior Research Fellowship. S.F.K. also thanks the CERN theory
division for its hospitality.


\end{document}